\documentclass[aps,prl,twocolumn,groupedaddress]{revtex4}
\usepackage[dvips]{graphics, color}

\begin{document}

\title{Momentum distribution of a Fermi gas of atoms in the BCS-BEC crossover}

\author{C. A. Regal$^{1}$}
\email[Email: ]{regal@jilau1.colorado.edu}
\author{M. Greiner$^{1}$}
\author{S. Giorgini$^{1,2}$}
\author{M. Holland$^{1}$}
\author{D. S. Jin$^{1}$}
\thanks{Quantum Physics Division, National Institute of Standards and Technology.}
\address{
$^{1}$JILA, National Institute of Standards and Technology and
University of Colorado,\\ and Department of Physics, University of
Colorado, Boulder, CO 80309-0440, USA\\ $^{2}$Dipartimento di
Fisica, Universit\`a di Trento and BEC-INFM, I-38050 Povo, Italy}

\date{\today}

\begin{abstract}
We observe dramatic changes in the atomic momentum distribution of
a Fermi gas in the region of the BCS-BEC crossover. We study the
shape of the momentum distribution and the kinetic energy as a
function of interaction strength. The momentum distributions are
compared to a mean-field crossover theory, and the kinetic energy
is compared to theories for the two weakly interacting limits. The
temperature dependence of the distribution is also presented.
\end{abstract}

\maketitle

Recent years have seen the emergence of an intriguing Fermi system
achieved with ultracold gases of $^{40}$K or $^6$Li atoms. With
these systems it is possible to widely tune the interatomic
interaction strength, represented dimensionlessly as $k_Fa$, where
$k_F$ is the Fermi wavevector and $a$ is the scattering length. Of
particular interest is the strongly interacting regime ($-1 <
1/k_Fa < 1$) where a crossover between BCS theory of
superconductivity and Bose-Einstein condensation (BEC) of
molecules occurs \cite{Randeriaetc,Holland2001a,Timmermans2001a}.
Experiments have shown that these Fermi systems cross a phase
transition as a function of temperature and display features of
the BCS-BEC crossover such as a pairing gap, $\Delta$, on order of
the Fermi energy, $E_F$. Experimental probes have been numerous
and include studies of molecule formation
\cite{Regal2003c,Cubizolles2003a,Jochim2003a,Strecker2003a,Hodby2005a},
thermodynamic properties
\cite{O'hara2002a,Regal2003b,Bourdel2003a,Bartenstein2004a,Bourdel2004a,Kinast2005a},
condensate formation in the crossover
\cite{Regal2004a,Zwierlein2004a}, collective excitations
\cite{Bartenstein2004b,Kinast2004a}, single-particle excitations
\cite{Chin2004a,Greiner2005a}, and vortices \cite{Zwierlein2005}.

\begin{figure} \begin{center}
\scalebox{.93}[.93]{\includegraphics{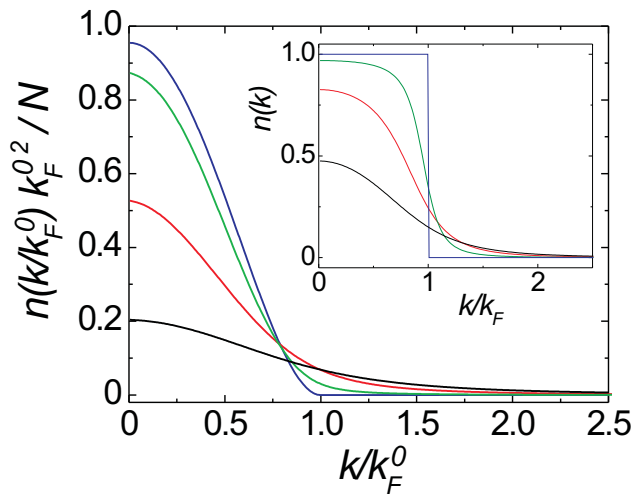}} \caption{(color
online) Theoretical column integrated momentum distributions of a
trapped Fermi gas $n(k)$ calculated from a mean-field theory
\cite{Viverit2004b}. $N$ is the total particle number, $k_F^0$ is
the non-interacting Fermi wavevector.  The normalization is given
by $2 \pi \int n(k) k dk = N$, and we note that the normalization
is strongly affected by the tail of the distribution. The lines
correspond to $a=0$ (blue), $1/k_F^0a=-0.66$ (green), $1/k_F^0a=0$
(red), and $1/k_F^0a=0.59$ (black). (inset) Corresponding
distributions for a homogeneous system.} \label{fig1}
\end{center}
\end{figure}

One classic phenomenon associated with pairing in a Fermi system
that has yet to be explored fully in atomic systems is a
broadening of the Fermi surface in momentum space (see for example
\cite{deGennes}). Figure \ref{fig1} (inset) shows the expected
momentum distribution of a homogeneous, zero temperature ($T=0$)
Fermi system. In the BCS limit ($1/k_Fa \rightarrow -\infty$) the
amount of broadening is small and associated with $\Delta$. As the
interaction increases this effect grows until at unitarity
($1/k_Fa=0$) the effect is on order of $E_F$, and in the BEC limit
($1/k_Fa \rightarrow \infty$) the momentum distribution becomes
the square of the Fourier transform of the molecule wavefunction.

In this Letter we present a measurement of the atom momentum
distribution of a trapped Fermi gas in the BCS-BEC crossover
regime. To perform these experiments we prepare a $^{40}$K gas
near a scattering resonance known as a Feshbach resonance, where
we can tune the s-wave interaction between fermions by varying the
magnetic field $B$ \cite{Regal2003a,Regal2003b}. To probe the
system we use the standard technique of time-of-flight expansion
followed by absorption imaging \cite{Anderson1995a}. To measure
the momentum distribution of atoms the gas must expand freely
without any interatomic interactions; to achieve this we quickly
change the scattering length to zero for the expansion. Bourdel
{\it et al.} pioneered this type of measurement using a gas of
$^6$Li atoms at $T/T_F \approx0.6$, where $T_F$ is the Fermi
temperature \cite{Bourdel2003a}. Here we report measurements down
to $T/T_F \approx 0.1$, where pairing becomes a significant effect
and condensates have been observed
\cite{Regal2004a,Zwierlein2004a}.

To understand what we expect for our trapped atomic system, we can
predict the atomic momentum distribution using a local density
approximation and the results for the homogeneous case. In the
trapped gas case, in addition to the local broadening of the
momentum distribution due to pairing, attractive interactions
compress the density profile and thereby enlarge the overall
momentum distribution. Figure \ref{fig1} shows an integrated
column density from the result of a mean-field calculation at
$T=0$ as described in Ref. \cite{Viverit2004b} \cite{montecarlo}.

In our experimental setup we create an ultracold $^{40}K$ gas
using previously described cooling techniques
\cite{DeMarco1999a,Regal2003b}.  The gas is prepared in a nearly
equal mixture of the spin-states $|f,m_f\rangle$ =
$|9/2,-9/2\rangle$ and $|9/2,-7/2\rangle$, where $f$ is the total
spin and $m_f$ the spin-projection quantum number.  The final
ultracold gas is held in an optical dipole trap formed at the
intersection of two gaussian laser beams. One beam is oriented
parallel to the force of gravity ($\hat{y}$) with a waist of
$w_y=$200 $\mu$m and the second beam is perpendicular to the first
($\hat{z}$) and has $w_z=$15 $\mu$m.

We first measured the atomic momentum distribution with our lowest
temperature Fermi gas.  We start with a weakly interacting gas at
$T \approx 0.12$ $T_F$ in a trap with a radial frequency of
$\nu_r=280$ Hz and an aspect ratio of $\nu_z/\nu_r=0.071$. We then
adiabatically increase the interaction strength by ramping the
magnetic field at a rate of (6.5 ms/G)$^{-1}$ to near a Feshbach
resonance located at $202.10 \pm 0.07$ G \cite{Regal2004a}. After
a delay of 1 ms, both dipole trap beams are switched off and
simultaneously a magnetic-field ramp to $a \approx 0$ ($B=209.6$
G) at a rate of (2 $\mu$s/G)$^{-1}$ is initiated. The rate of this
magnetic-field ramp is designed to be fast compared to typical
many-body timescales as determined by $\frac{h}{E_F}=90$ $\mu$s.
The cloud is allowed to freely expand for 12.2 ms, and then an
absorption image is taken. The imaging beam propagates along
$\hat{z}$ and selectively probes the $|9/2,-9/2\rangle$ state
\cite{Regal2003b}.

\begin{figure} \begin{center}
\scalebox{.93}[.93]{\includegraphics{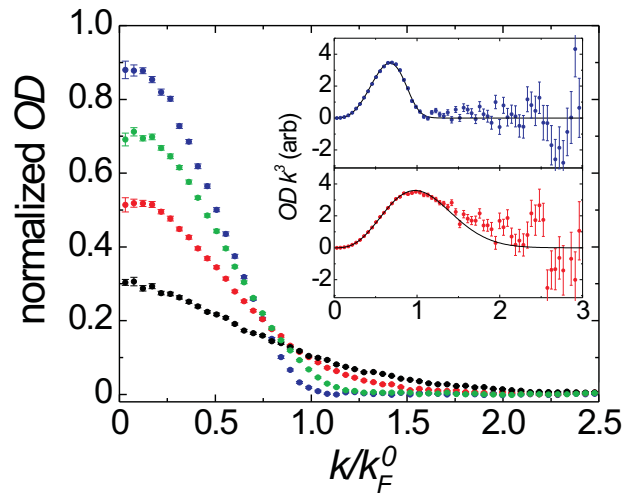}} \caption{(color
online) Experimental, azimuthally averaged, momentum distributions
of a trapped Fermi gas normalized such that the area under the
curves is the same as in Fig. \ref{fig1}. The curves correspond to
$1/k_F^0a=-71$ (blue), -0.66 (green), 0 (red), and 0.59 (black).
Error bars represent the standard deviation of the mean of
averaged pixels. (inset) Curves for $1/k_F^0a=-71$ (top) and 0
(bottom) weighted by $k^3$. The lines are the results of a fit to
Eqn. \ref{eqn1}.} \label{fig2}
\end{center}
\end{figure}

Samples of these absorption images, azimuthally averaged, are
shown in Fig. \ref{fig2} for various values of $1/k_F^0a$, where
the superscript $0$ indicates a quantity that was measured in the
weakly interacting regime.  We observe a dramatic change in the
distribution as predicted in Fig. \ref{fig1}.  Some precautions
need to be taken in quantitative comparison of Figs. \ref{fig1}
and \ref{fig2}.  First, the magnetic-field ramp to the Feshbach
resonance, while adiabatic with respect to most time scales, is
not fully adiabatic with respect to the axial trap period. Second,
in the experiment an adiabatic field ramp keeps the entropy of the
gas, not $T/T_F$, constant. However, we expect the resulting
change in $T/T_F$ to have a minimal effect on the distribution for
$1/k_F^0a<0$ \cite{Chen2005a}.  Third, the theory assumes $T=0$
and does not include the Hartree term, thus underestimating the
broadening on the BCS side compared to a full theory
\cite{Astrakharchik2004a}.

It is natural now to consider extracting the kinetic energy from
the momentum distribution. While the momentum distribution should
be universal for small momenta, for large momenta it is influenced
by details of the interatomic scattering potential. In the extreme
case of a delta potential, which we used for the calculation in
Fig. 1, the momentum distribution has a tail with a $1/k^4$
dependence, giving rise to a divergence of the kinetic energy. In
the experiment we avoid a dependence of the measured kinetic
energy on details of the interatomic potential because our
magnetic-field ramp is never fast enough to access features on
order of the interaction length of the Van der Waals potential,
$r_0$ $\approx 60$ $a_0$ for $^{40}$K \cite{Gribakin1993a}. Thus,
the results presented in this Letter represent a universal
quantity, independent of the details of the interatomic potential.
Although universal in this sense, the measured kinetic energy is
intrinsically dependent on the dynamics of the magnetic-field
ramp, with faster ramps corresponding to higher measured energies.

To obtain the kinetic energy from the experimental data exactly we
would need to take the second moment of the distribution, which is
proportional to $\sum k^3 OD/\sum k OD$, where $OD$ is the optical
depth. As illustrated in Fig. \ref{fig2} (inset) this is difficult
due to the decreased signal-to-noise ratio for large $k$. Thus,
our approach will be to apply a 2D surface fit to the image and
extract an energy from the fitted function. In the limit of weak
interactions the appropriate function is that for an ideal,
harmonically trapped Fermi gas. This is
\begin{equation} OD(x,y)=pk \hspace{0.07cm} g_2 (-\zeta  \hspace{0.05cm} e^{-\frac{x^2}{2
\sigma_x^2}-\frac{y^2}{2 \sigma_y^2}})/g_2 (-\zeta) \label{eqn1}
\end{equation} where
$g_n(x)=\sum\limits_{k=1}^{\infty}\frac{x^k}{k^n}$, $\zeta$ is the
fugacity, $\sigma_{x,y}^2$ are proportional to the Fermi gas
temperature, and $pk$ is the maximum OD. Assuming isotropic
expansion in all three dimensions the kinetic energy per particle
is given by
\begin{equation} E_{kin} = \frac{3}{2} \frac{m \sigma_x
\sigma_y}{t^2} \frac{g_4(-\zeta)}{g_3(-\zeta)} \label{eqn2}
\end{equation} where $m$ is the mass of $^{40}$K and $t$ is the
expansion time. Empirically, we find Eqn. \ref{eqn1} fits
reasonably well to data throughout the crossover, as illustrated
in Fig. \ref{fig2} (inset).

\begin{figure} \begin{center}
\scalebox{.9}[.9]{\includegraphics{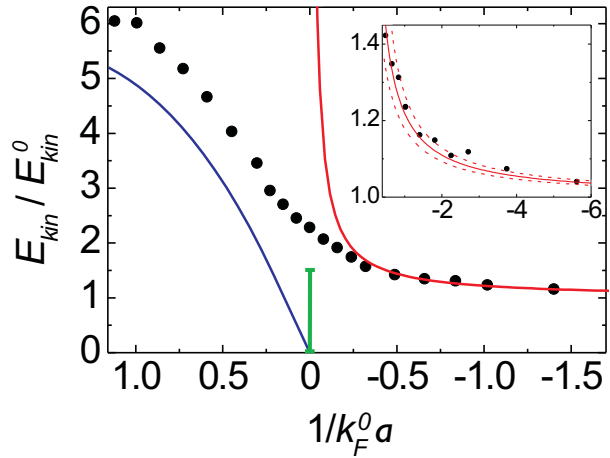}} \caption{(color
online) The points ($\bullet$) show the measured energy $E_{kin}$
 normalized to $E_{kin}^0=0.25$
$k_b$ $\mu$K ($E_{kin}^0=\frac{3}{8} E_F$ for a harmonically
trapped gas at $T=0$). The red line is the expected energy ratio
from a calculation only valid in the weakly interacting regime
($1/k_Fa<-1$). In the strongly interacting regime pairing due to
many-body effects strongly increases $E_{kin}$. The green bar
represents the expected value of $E_{kin}/E_{kin}^0$ at unitarity
just due to density profile rescaling (see text). In the molecule
limit ($1/k_F^0a > 1$) we calculate the expected energy for an
isolated molecule (blue line). (inset) A focus on the weakly
interacting regime with the same axes definitions as the main
graph. The dashed lines show the uncertainty in the calculation
based upon the uncertainty in the Feshbach resonance parameters
\cite{Regal2003a,Regal2004a}.} \label{fig3}
\end{center}
\end{figure}

Figure \ref{fig3} shows the result of extracting $E_{kin}$ as a
function of $1/k_F^0a$; we see that $E_{kin}$ more than doubles
between the non-interacting regime and unitarity. We have checked
that heating and loss due to inelastic processes are negligible up
to $1/k_F a \sim 0$. To do this we performed an experiment in
which we adiabatically approach the Feshbach resonance at rate of
(6 ms/G)$^{-1}$, wait 1 ms, and then ramp back at the same slow
rate to the weakly interacting regime. If we start with a cloud
initially at $T/T_F=0.10$, $T/T_F$ upon return increases by less
than $10\%$ for a ramp to $1/k_F a=0$ (yet by $80\%$ for a ramp to
$1/k_F^0 a=0.5$).

\begin{figure} \begin{center}
\scalebox{.92}[.92]{\includegraphics{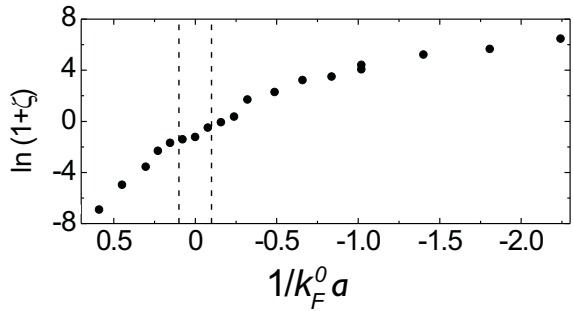}} \caption{Shape of
the momentum distribution as described through the parameter
$\zeta$ (Eqn. \ref{eqn1}). ln$(1+\zeta)=0$ corresponds to a
gaussian distribution, and for an ideal Fermi gas
ln$(1+\zeta)^{-1} \approx T/T_F$ in the limit of low $T/T_F$.  The
dashed lines show the uncertainty in the Feshbach resonance
position  \cite{Regal2004a}.} \label{fig4}
\end{center}
\end{figure}

Using the fitting function of Eqn. \ref{eqn1} we can also extract
information about the shape of the distribution through the
parameter $\zeta$. Since $\zeta$ can range from -1 to $\infty$ it
is convenient to plot the quantity ln$(1+\zeta)$ (Fig.
\ref{fig4}). We find that the shape evolves smoothly from that of
an ideal Fermi gas at $T/T_F \sim 0.1$ in the weakly interacting
regime, to a gaussian near unitarity, and to a shape more peaked
than a gaussian in the BEC regime.  These qualitative features are
predicted by the mean-field calculation of the distributions in
Fig. \ref{fig1}.

As mentioned earlier $E_{kin}$ of a trapped gas is affected both
by the broadening due to pairing (Fig. \ref{fig1} (inset)) and by
changes in the trapped gas density profile.  In the BCS limit, the
broadening due to pairing scales with $e^{-\pi/2k_F |a|}$ and is
thus exponentially small compared to density profile changes,
which scale linearly with $k_F |a|$. In this limit we can
calculate $E_{kin} / E_{kin}^0$ using a mean-field calculation in
the normal state \cite{Vichi1999a} to find, to lowest order in
$k_F^0|a|$, $E_{kin} / E_{kin}^0=\frac{2048}{945\pi^2}
k_F^0|a|+1$. We plot this result in Fig. \ref{fig3} (inset) and
find good agreement for the weakly interacting regime
($1/k_F^0|a|>1$). In the crossover regime where the pairs are more
tightly bound, pairing provides a significant contribution to the
change in the momentum distribution. At unitarity a full Monte
Carlo calculation predicts the radius of the Fermi gas density
profile to become $(1+\beta)^{1/4} R_0=0.81 R_0$, where $R_0$ is
the Thomas-Fermi radius of a non-interacting Fermi gas
\cite{Astrakharchik2004a}. Just this rescaling would result in $
E_{kin}/E_{kin}^0=1.54$ (green bar in Fig. \ref{fig3}). Thus, at
unitarity, pairing effects on the momentum distribution must
account for a large fraction of the measured value of
$E_{kin}/E_{kin}^0=2.3 \pm 0.3$ (Fig. \ref{fig3}) and all of the
observed change in distribution shape (Fig. \ref{fig4}).

In the BEC limit we expect the measured energy to be that of an
isolated diatomic molecule after dissociation by the
magnetic-field ramp.  Provided the scattering length associated
with the initial molecular state, $a(t=0)$, is much larger than
$r_0 \approx 60$ $a_0$, the wave function for the molecule is
given by $\psi=e^{-r/a(t=0)}/r$ where $r$ is the internuclear
separation. We can calculate the measured energy from the solution
of the Schr\"odinger equation with a time-dependent boundary
condition on the two-particle wavefunction
$\left.\frac{d\log(r\psi)}{dr}\right|_{r=0}=-\frac{1}{a(t)}$,
where $a(t)$ is the scattering length fixed by the magnetic field
at time $t$. In Fig. \ref{fig3} we show the result of this
calculation for a pure gas of molecules and a (2 $\mu$s/G)$^{-1}$
ramp rate. We find reasonable agreement considering that there is
a large systematic uncertainty in the theory curve due to the
experimental uncertainty in the magnetic-field ramp rate and that
this two-body theory should match the data only in the BEC limit
($1/k_Fa \gg 1$).

\begin{figure} \begin{center}
\scalebox{1}[1]{\includegraphics{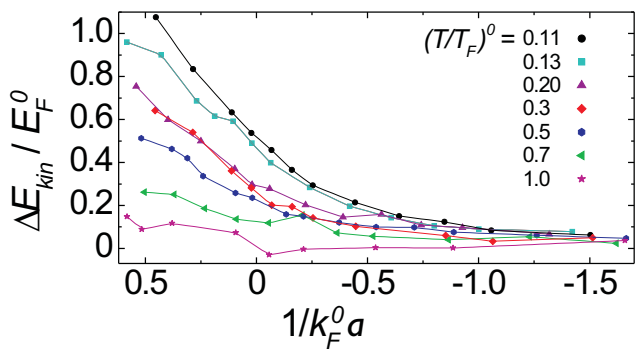}} \caption{(color
online) Temperature dependence of $\Delta
E_{kin}=E_{kin}-E_{kin}^0$ normalized to the Fermi energy at
$a=0$. $(T/T_F)^0$ is the temperature of the non-interacting gas
\cite{tempnote}. For the coldest dataset (black) the peak density,
for atoms in one of the two spin states, in the weakly interacting
regime is $n_{pk}^0 = 1.4 \times 10^{13}$ cm$^{-3}$ and
$E_F^0=0.56$ $\mu$K. For the hottest dataset (magenta) $n_{pk}^0$
decreases to $6 \times 10^{12}$ cm$^{-3}$ and $E_F^0=0.79$
$\mu$K.} \label{fig5}
\end{center}
\end{figure}

We have also studied the dependence of the momentum distribution
on $(T/T_F)^0$.  To vary the temperature of our gas, we recompress
the optical dipole trap after evaporation and parametrically heat
the cloud \cite{Hodby2005a}. The experimental sequence for
measuring the momentum distribution is the same as above except
the ramp rate to $a=0$ for expansion was $\sim$ (8
$\mu$s/G)$^{-1}$. Figure \ref{fig5} shows the measured kinetic
energy change $\Delta E_{kin}=E_{kin}-E_{kin}^0$. On the BEC side
of the resonance ($1/k_Fa>0$), $\Delta E_{kin}$ decreases
dramatically with $(T/T_F)^0$. Because $\Delta E_{kin}$ should be
proportional to the molecule fraction, this result is closely
related to the recent observation that the molecule conversion
efficiency scales with $T/T_F$ \cite{Hodby2005a}.  In the strongly
interacting regime we also observe a decrease in $\Delta E_{kin}$
with increasing $(T/T_F)^0$.  Here the temperature dependence of
$\Delta E_{kin}$ is consistent with the expectation that the
changes in the kinetic energy are caused by pairing and not
coherence \cite{Regal2004a,Chin2004a,Chen2004a}.

We have found that the momentum distribution of a Fermi gas
provides a wealth of information on crossover physics.  The
measurement is a probe of pairing in the strongly interacting
regime, and it provides a universal thermodynamic quantity that is
complementary to previously measured energies in the BCS-BEC
crossover
\cite{O'hara2002a,Regal2003b,Bourdel2003a,Bartenstein2004a,Bourdel2004a}.
This work also provides a starting point for future experiments
seeking to probe pair correlations in the momentum distribution
using measurements of atom shot noise \cite{Greiner2005b}.

We thank Q. Chen and K. Levin for valuable discussions and J. T.
Stewart for experimental assistance. This work was supported by
NSF, NIST, and NASA. C. A. R. acknowledges support from the Hertz
Foundation.


\begin{thebibliography}{10}
\bibitem{tempnote}
$(T/T_F)^0$ is measured by fitting to Eqn. 1 the momentum
distributions
  measured near $a=0$, probing along both $\hat{x}$ and $\hat{z}$. Above
  $T/T_F=0.45$ we fix $\zeta$ to zero and use a calculated correction factor to
  determine $T/T_F$.
\end{thebibliography}

\end{document}